\documentclass[amsmath,nobibnotes,superscriptadress,amssymb,aps,pra,showpacs,superscriptaddress, twocolumn]{revtex4-1}
\usepackage{amsmath,amsfonts,amssymb,amsthm,graphics,graphicx,epsfig,bbm}
\usepackage[colorlinks=true,citecolor=green,linkcolor=blue,urlcolor=magenta]{hyperref}
\usepackage[usenames]{color}
\usepackage{footnote}
\usepackage{graphicx}
\usepackage{subfigure}
\usepackage{amsmath}
\usepackage{epsfig}
\usepackage{dcolumn}
\usepackage{lmodern}
\usepackage{enumitem}
\usepackage{bm}
\usepackage{color}
\usepackage{times}
\usepackage{epstopdf}
\usepackage{amssymb}
\usepackage{amstext}
\usepackage{latexsym}
\usepackage{hyperref}
\usepackage{amsfonts}
\usepackage{psfrag}
\usepackage{soul,xcolor}
\usepackage[normalem]{ulem}
\usepackage{dsfont}
\usepackage{txfonts}
\usepackage{footnote}
\usepackage{multirow}
\usepackage{appendix}
\usepackage{mathtools}
\usepackage{xspace}

\renewcommand{\Re}{\mathrm{Re}}
\renewcommand{\Im}{\mathrm{Im}}

\usepackage{mathtools}

\begin{document}
\setstcolor{red}
\newtheorem{Proposition}{Proposition}[section]	
\title{Probing Dynamical Sensitivity of a Non-KAM System Through Out-of-Time-Order Correlators} 
\date{\today}

\author{Naga Dileep Varikuti}
\email{vndileep@physics.iitm.ac.in}
\affiliation{Department of Physics, Indian Institute of Technology Madras, Chennai, India, 600036}
\affiliation{Center for Quantum Information, Communication and Computation,
Indian Institute of Technology Madras, Chennai, India 600036}
\author{Abinash Sahu}
\affiliation{Department of Physics, Indian Institute of Technology Madras, Chennai, India, 600036}
\affiliation{Center for Quantum Information, Communication and Computation,
Indian Institute of Technology Madras, Chennai, India 600036}
\author{Arul Lakshminarayan}
\affiliation{Department of Physics, Indian Institute of Technology Madras, Chennai, India, 600036}
\affiliation{Center for Quantum Information, Communication and Computation,
Indian Institute of Technology Madras, Chennai, India 600036}
\author{Vaibhav Madhok}
\affiliation{Department of Physics, Indian Institute of Technology Madras, Chennai, India, 600036}
\affiliation{Center for Quantum Information, Communication and Computation,
Indian Institute of Technology Madras, Chennai, India 600036}

\begin{abstract}
Non-KAM (Kolmogorov-Arnold-Moser) systems, when perturbed by weak time-dependent fields, offer a fast route to classical chaos through an abrupt breaking of invariant phase space tori. In this work, we employ out-of-time-order correlators (OTOCs) to study the dynamical sensitivity of a perturbed non-KAM system in the quantum limit as the parameter that characterizes the \textit{resonance} condition is slowly varied. For this purpose, we consider a quantized kicked harmonic oscillator (KHO) model, which displays stochastic webs resembling Arnold's diffusion that facilitate large-scale diffusion in the phase space. Although the Lyapunov exponent of the KHO at resonances remains close to zero in the weak perturbative regime, making the system weakly chaotic in the conventional sense, the classical phase space undergoes significant structural changes. Motivated by this, we study the OTOCs when the system is in resonance and contrast the results with the non-resonant case. At resonances, we observe that the long-time dynamics of the OTOCs are sensitive to these structural changes, where they grow quadratically as opposed to linear or stagnant growth at non-resonances. On the other hand, our findings suggest that the short-time dynamics remain relatively more stable and show the exponential growth found in the literature for unstable fixed points. The numerical results are backed by analytical expressions derived for a few special cases. We will then extend our findings concerning the non-resonant cases to a broad class of near-integrable KAM systems.          
\end{abstract}

\maketitle

\newtheorem{theorem}{Theorem}[section]
\newtheorem{corollary}{Corollary}[theorem]
\newtheorem{lemma}[theorem]{Lemma}
\def\endproof{\hfill$\blacksquare$}
\section{introduction}
\label{section-1}
Quantum chaos is the study of quantum systems whose classical counterparts are chaotic. An overwhelming majority of such studies have considered Hamiltonian chaos in classical systems, where the celebrated Kolmogorov-Arnold-Moser (KAM) theorem is applicable and studied the signatures of classical chaos in the quantum domain. The KAM theorem states that if an integrable Hamiltonian system, which admits as many independent constants of motion as the total system degrees of freedom, is subjected to a weak generic perturbation, most invariant tori in the phase space will persist with slight deformations \cite{arnold2009proof, kolmogorov1954conservation, moser1962invariant, moser1967convergent}. The chaos in such systems manifests through the gradual destruction of the invariant tori. However, the validity of the KAM theorem rests upon a few fundamental assumptions. For example, it presupposes that the unperturbed Hamiltonian is non-degenerate, meaning that when expressed in the action-angle variables, the Hamiltonian takes the form of a nonlinear function involving only the action variables while the angle variables remain cyclic. In addition, the characteristic frequency ratios must be sufficiently irrational for the phase space tori to survive the perturbations. Upon failing to meet these assumptions, the tori will likely break immediately at any arbitrary perturbation, leading to the emergence of widespread chaos \cite{poschel2009lecture}. Most realistic physical systems satisfy the KAM conditions. There is, however, a family of non-KAM systems that do not follow the usual KAM route to chaos.

In this work, we ask the following question: How sensitive is the information scrambling to the perturbations in a quantum system whose classical counterpart is non-KAM? We tackle this question by studying the scrambling at classical resonances, which are the salient features of a non-KAM system. At the resonances, the non-KAM systems display large-scale structural changes in the presence of perturbations. In the classical phase space, the resonances are generally associated with breaking the invariant phase space tori via the creation of stable and unstable phase space manifolds. Such a mechanism results in diffusive chaos in the phase space even when the perturbation is arbitrarily small. Thus, the non-KAM systems show high sensitivity to the small changes in the system parameters at the resonances. Earlier works exemplified the dynamics of non-KAM systems by studying the systems that transit from being discontinuous to continuous depending upon the values of the appropriate parameters \cite{sankaranarayanan2001quantum, sankaranarayanan2001chaos, paul2016barrier, santhanam2022quantum}. We instead focus on a system that exhibits non-KAM behavior as a consequence of the classical degeneracy. With this objective in mind, we adopt the kicked harmonic oscillator (KHO) model as a paradigm to study information scrambling. We use out-of-time-ordered correlators (OTOCs) to diagnose and investigate the sensitivity of scrambling at the resonances and the non-resonances of the quantum KHO.

The OTOCs were first introduced in the context of superconductivity \cite{larkin} and have been recently revived in the literature to study information scrambling in many-body quantum systems \cite{ope2, ope1, ope4, ope5, lin2018out, shukla2022out}, quantum chaos \cite{chaos1, pawan, rozenbaum2017lyapunov, seshadri2018tripartite, lakshminarayan2019out, shenker2, moudgalya2019operator, omanakuttan2019out, manybody2, chaos2, prakash2020scrambling, prakash2019out, varikuti2022out, markovic2022detecting}, many-body localization \cite{manybody3, manybody4, manybody1, huang2017out} and holographic systems\cite{shock1, shenker3}. Given two operators $A$ and $B$, the out-of-time-ordered commutator function in an arbitrary quantum state $|\psi\rangle$ is given by
\begin{eqnarray}\label{commutator}
C_{|\psi\rangle,\thinspace AB}(t)=\langle\psi| \left[A(t), B\right]^{\dagger}\left[A(t), B\right]|\psi\rangle,
\end{eqnarray}
where $A(t)=\hat{U}^{\dagger}(t)A\hat{U}(t)$ is the Heisenberg evolution of $A$ governed by the Hamiltonian evolution of the system. When expanded, the function $C_{|\psi\rangle,\thinspace AB}(t)$ contains two-point and four-point correlators. Since the time ordering in these four-point correlators is non-sequential, they are usually referred to as the OTOCs. The behavior of $C_{AB}(t)$ depends predominantly on the four-point correlators. Hence, the terms OTOC and the commutator function are often used interchangeably to denote the same quantity $C_{|\psi\rangle, AB}(t)$. 

To understand how the OTOCs diagnose chaos, consider the phase space operators $\hat{X}$ and $\hat{P}$ in the semiclassical limit ($\hbar\rightarrow 0$), where the Poisson brackets replace the commutators. It can be readily seen that $\{ X(t), P \}^2=(\delta X(t)/\delta X(0))^2\sim e^{2\lambda t}$, where $\lambda$ denotes Lyapunov exponent of the system under consideration, which is positive for chaotic systems. The correspondence principle then establishes that the OTOCs of a quantum system, whose classical limit is chaotic, grow exponentially until a time known as Ehrenfest’s time $t_{\text{EF}}$ that depends on the dynamics of the system \cite{schubert2012wave, rozenbaum2020early, jalabert2018semiclassical, chen2018operator}. For the single particle chaotic systems, $t_{\text{EF}}$ scales logarithmically with the effective Planck constant and inversely with the corresponding classical Lyapunov exponent --- $t_{\text{EF}}\sim \ln(1/\hbar_{\text{eff}})/\lambda$. For $t>t_{\text{EF}}$, the correspondence breaks down due to the non-trivial $\hbar$ corrections arising from the phase space spreading of the initially localized wave packets. While many recent works have revealed that the early-time growth rate of OTOCs correlates well with the classical Lyapunov exponent for the chaotic systems, it is, however, worthwhile to note that the exponential growth may not always represent true chaos in the system \cite{pappalardi2018scrambling, hashimoto2020exponential, pilatowsky2020positive, pilatowsky2020positive, xu2020does, hummel2019reversible, steinhuber2023dynamical}. Nonetheless, by carefully treating the singular points of the system, one can show that the OTOCs continue to serve as a reliable diagnostic of chaos \cite{wang2022statistical, wang2021quantum}.  

Recently, the OTOCs have been found to show intriguing connections with other probes of quantum chaos such as tripartite mutual information \cite{pawan}, operator entanglement \cite{styliaris2021information, zanardi2021information}, quantum coherence \cite{anand2021quantum} and Loschmidt echo \cite{yan2020information} to name a few. Moreover, the OTOCs have been investigated in the deep quantum regime and observed that the signatures of short-time exponential growth can still be found in such systems \cite{sreeram2021out}. Also, see Ref. \cite{pg2022witnessing} for an interesting comparison of the OTOCs with observational entropy, a recently introduced quantity to study the thermalization of closed quantum systems \cite{vsafranek2019quantum, vsafranek2019quantum1}. 

This paper is structured as follows. In Sec. \ref{section-2} we review some basic features of the KHO model, including resonances and non-resonances in both classical and quantum domains. We analyze the behavior of OTOCs in Sec. \ref{section-3} with a special emphasis given to the short-time dynamics in Sec. \ref{resonanceOTOC}. Thereafter, we focus on the asymptotic time dynamics of the OTOCs in Sec. \ref{non-resonanceOTOC} and show how the OTOCs distinguish the resonances from the non-resonances. In Sec. \ref{analtytical} we analytically derive the OTOCs for a few special cases of the quantum KHO model. Then, in Sec. \ref{OTOCXP} we provide a brief overview of the OTOCs for the phase space operators. We finally conclude this text in Sec. \ref{section-4} with a few remarks on the relevance of this work to the stability of quantum simulators.

\section{Model: Kicked harmonic oscillator}
\label{section-2}
We consider the harmonic oscillator model with a natural frequency $\omega$, subjected to periodic kicks by a nonlinear position-dependent field, having the following Hamiltonian \cite{chernikov1989symmetry, billam2009quantum, berman1991problem, kells2005quantum, afanasiev1990width, reichl2021transition, gardiner1997quantum, rechester1980calculation, ichikawa1987stochastic, ishizaki1991anomalous, daly1994classical, borgonovi1995translational, engel2007quantum}:
\begin{eqnarray}
H=\dfrac{P^2}{2m}+\dfrac{1}{2}m\omega^2X^2+ K\cos(kX)\sum_{n=-\infty}^{\infty}\delta\left(t-n\tau\right),
\end{eqnarray}
where $X$ is the position, $P$ is the momentum, $m$ is the mass of the oscillator and $k$ is the wave vector. The strength of the kicking is denoted by $K$. The time interval between two consecutive kicks is given by $\tau$. For simplicity, throughout the paper, we will take $m=k=1$. The system is parity invariant, i.e., $H(P, X)=H(-P, -X)$. 

In the action-angle coordinates, the Hamiltonian of the harmonic oscillator ($H_0$) scales linearly with the action coordinate --- the canonical transformation to the action-angle variables as given by $(X, P)= (\sqrt{2I/\omega}\cos\theta, \sqrt{2I\omega}\sin\theta)$ yields $H_0=\omega I$. Since $H_0$ is linear in $I$, the characteristic frequency of the phase space tori $(\partial H_0/\partial I=\omega)$ turns out to be independent of $I$, which violates one of the KAM assumptions that the unperturbed Hamiltonian must be non-linear in $I$ (the non-degeneracy condition). Hence, the harmonic oscillator is classified as non-KAM integrable. In the following, we briefly discuss the dynamical aspects of the KHO model in both the classical and quantum limits. 

\subsection{Classical dynamics}
\begin{figure}
\includegraphics[scale=0.35]{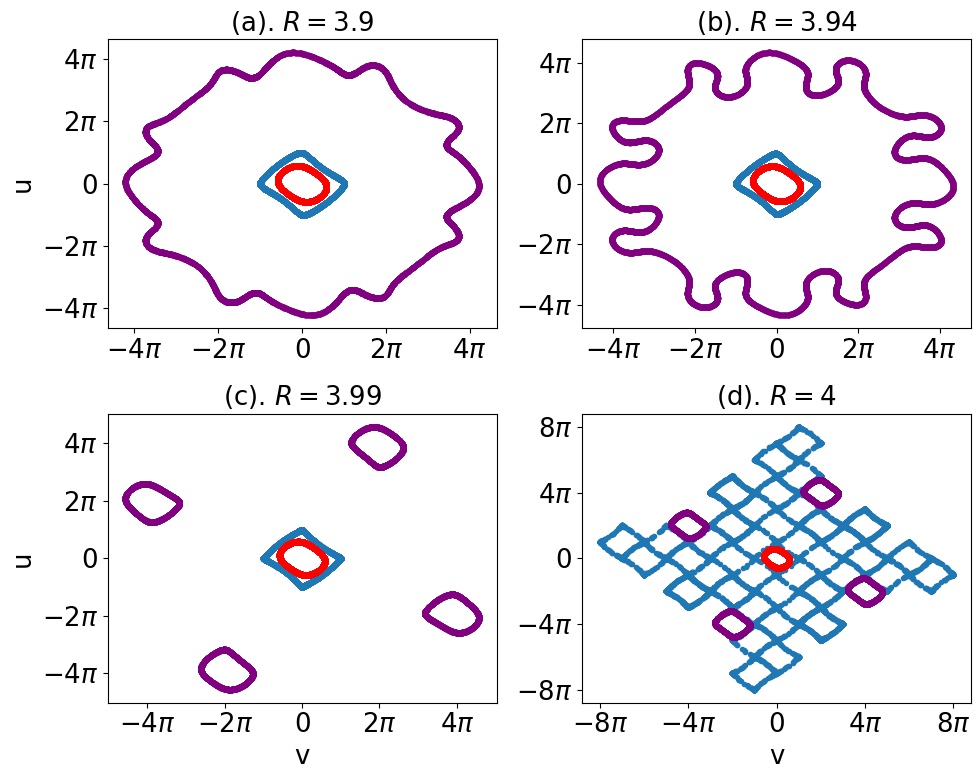}
\caption{\label{fig:poincare}The figure illustrates the classical phase space trajectories of the KHO model for three randomly chosen initial conditions in the vicinity of $R=4$, and each evolved for $10^4$ time steps. Here, we set $K=1$ and $\tau=1$. When $R=4$, for the given set of parameters, the phase space mainly constitutes three regions: the stochastic web, period four islands, and a period one regular island surrounding the origin. Each initial condition in the figure corresponds to one of these regions. When the system is fully non-resonant, the phase space is mostly regular. With $R$ varying from $3.9$ to $4$ in the sequential order depicted in panels (a)-(d), the trajectories become increasingly distorted until they completely break apart at the resonance. The periodic islands feature stable periodic points at their centers and unstable periodic points on their exterior. The boundaries of these islands collectively constitute the stochastic web that facilitates large-scale phase space diffusion.}
\end{figure}
In the classical limit, the dynamics of the KHO model can be visualized by the following two-dimensional map:
\begin{eqnarray}\label{dynmap}
u_{n+1} &=&(u_n+\epsilon\sin v_n)\cos(\omega\tau) +v_n \sin(\omega\tau),\nonumber\\
v_{n+1} &=&-(u_n+\epsilon\sin v_n)\sin(\omega\tau) +v_n \cos(\omega\tau),
\end{eqnarray}
where $u=P/\omega$, and $v=X$ denote scaled momentum and position variables, respectively and $\epsilon=K/\omega$ is the effective strength of perturbation. In the remainder of this paper, we adopt the notation $\omega\tau=2\pi/R$. The system is considered classically resonant (or simply in resonance) whenever $R$ assumes integer values. For non-integer $R$, the system is non-resonant. The KHO map can be physically realized through the motion of a charged particle in a constant magnetic field, with the particle being kicked by the wave packets of an electric field that propagates perpendicular to the direction of the magnetic field \cite{zaslavsky2007physics}.

Under the rescaling of the variables $(X, P/\omega)\rightarrow (v, u)$, the unperturbed harmonic oscillator is rotationally invariant in the phase space. Besides, the kicking potential, $\cos X$, is invariant under the translations along the $X$-direction by integer multiples of $2\pi$. Thus, when $K$ is non-zero, there will be a natural competition between the translational and rotational symmetries of the phase space. This competition becomes more prominent for $R\in R_c\equiv\{1, 2, 3, 4, 6\}$ \cite{chernikov1989symmetry, afanasiev1990width}. The system is exactly solvable for $R=\{1, 2\}$. In the remaining cases, the phase space consists of periodic stochastic webs. These webs display translation and rotational invariance in the phase space. In particular, the cell structure of these webs closely resembles tessellations. For $R=4$, the web appears as a square lattice, and for $R=3$ and $6$, it is a Kagome lattice with hexagonal symmetry. In this work, we focus on the vicinity of $R=4$ for the studies of information scrambling. 

The stochastic webs resemble Arnold's diffusion in systems with more than two degrees of freedom. Their thickness is exponentially small in $\epsilon$ \cite{afanasiev1990width}. These webs arise from the exteriors of the periodic islands and mainly consist of the separatrices of the KHO. Any trajectory that sets out on the separatrices will eventually diffuse towards the infinity --- $\langle r_n\rangle\sim \epsilon\sqrt{n}$, where $r_n=\omega\sqrt{P^2+X^2}$ is the distance traversed by an average phase space trajectory after $n$ time steps. As a result, the mean energy of the system grows linearly at any finite perturbation --- $\langle E_n\rangle\sim\epsilon^2 n$. The diffusion, however, is suppressed for weak perturbations when $R$ takes non-integer values. In the latter case, the diffusion coefficient remains close to zero for small perturbations \cite{kells2005quantum}. Nevertheless, the differences between the resonances and the non-resonances become less apparent as the perturbation increases. Figure \ref{fig:poincare} shows the phase space trajectories of the KHO system for a few randomly chosen initial conditions in the vicinity of $R=4$ for $K=1$. The corresponding separatrix equation is given by $v=\pm(u+\pi)+2l\pi$, $l\in\mathbb{Z}$ \cite{afanasiev1990width}. The phase space is regular with distorted circles when the system is non-resonant. However, it can be seen from the figure that the trajectories get increasingly deformed as $R$ approaches $4$. At $R=4$, the phase space undergoes significant changes due to the creation of period-$4$ orbits. Such behavior has applications in the chaotic electron transport in semiconductor superlattices \cite{fromhold2001effects, fromhold2004chaotic}.
\subsection{Quantum dynamics}
The existence of stochastic webs in the classical phase space can have far-reaching consequences on the corresponding quantum dynamics, which we briefly discuss here to set the ground for the OTOC analysis in the next section. As the system is being kicked at periodic intervals of time, the time-evolution is given by the following Floquet operator:
\begin{equation}
\hat{U}_\tau=\exp\left\{-\dfrac{2\pi i}{R} \hat{a}^\dagger \hat{a}\right\}\exp\left\{-\dfrac{iK}{\hbar}\cos\hat{X} \right\},
\end{equation}
where $\hat{a}$ and $\hat{a}^{\dagger}$ are the annihilation and creation operators corresponding to the particle trapped in the harmonic potential, respectively. The position operator is $\hat{X}=\sqrt{\hbar/2\omega}(\hat{a}+\hat{a}^{\dagger})$. The irrelevant global phase $e^{-i\pi /R}$ is ignored. The quantum chaos in the KHO model has been extensively studied over many years \cite{berman1991problem, shepelyansky1992quantum, daly1994classical, kells2005quantum, daly1996non, engel2007quantum, kells2004dynamical}. Experimental proposals have also been put forth to realize the dynamics of quantum KHO using ion traps and Bose-Einstein condensates \cite{gardiner1997quantum, carvalho2004web, gardiner2000nonlinear, duffy2004nonlinear, billam2009quantum}.

Recall that the classical KHO displays translational invariance whenever $R\in R_{c}$, which in the quantum limit, translates into the existence of commuting groups of translations. In particular, under the translation invariance, the $R$-th powers of $\hat{U}_{\tau}$ commute with either one or two parameter groups of translations or displacement operators \cite{borgonovi1995translational}. As a result, the system admits extended Floquet states in the phase space, leading to an unbounded growth of the mean energy $\langle(a^\dagger a)(t)\rangle$. Moreover, these states also facilitate the dynamical tunneling of the localized coherent states \cite{carvalho2004web}. As we shall see later, the translation invariance is also crucial to obtaining certain analytical expressions of the OTOCs, which are otherwise intractable.

For the non-integer $R$ values, previous studies argued that quantum localization takes place, and energy growth will be stopped after some time \cite{borgonovi1995translational}. More specifically, when $R$ is an irrational number, the quantum KHO model can be mapped to a tight-binding approximation in the limit of $K\lesssim \hbar\pi$, which explains the localization of the quantum dynamics \cite{frasca1997quantum, kells2005quantum}. The effects of localization are also prevalent in the scrambling dynamics. We will explore this in more detail in the coming sections.

\section{ Scrambling in quantum KHO model}
\label{section-3}
In this section, we address the central goal of our paper, which is to contrast the dynamics of information scrambling at resonances with that of non-resonances of the quantum KHO. The OTOCs are natural candidates to quantify the scrambling. While the OTOCs have been extensively studied in finite-dimensional systems, including those with time dependence \cite{prakash2020scrambling, sreeram2021out, borgonovi2019timescales, borgonovi2019timescales, shen2017out, zamani2022out}, the studies on continuous variable systems have primarily focused on time independent settings \cite{zhuang2019scrambling, hashimoto2020exponential}. However, the system considered in this paper is both continuous variable and time-dependent, leading to possible unbounded orbits in phase space and consequently to unbounded growth of OTOCs.

To study the OTOCs in the continuous variable systems, an appropriate choice of initial operators would be the canonical pair of position ($\hat{X}$) and momentum ($\hat{P}$) operators. However, for the reasons that become clear later, we instead consider the bosonic ladder operators $\hat{a}$ and $\hat{a}^{\dagger}$ as the initial operators. For an arbitrary state $|\psi\rangle$, we are interested in evaluating the quantity
\begin{eqnarray}\label{sf}
C_{|\psi\rangle,\thinspace \hat{a}\hat{a}^{\dagger}}(t)=\langle\psi| [\hat{a}(t), \hat{a}^\dagger]^\dagger [\hat{a}(t), \hat{a}^\dagger]|\psi\rangle,  
\end{eqnarray}
where $\hat{a}(t)=\hat{U}_{\tau}^{\dagger t} \hat{a} \hat{U}_{\tau}^t$ is the Heisenberg evolution of $\hat{a}$ under the dynamics of KHO and $t$ is the total number of time steps. Before analysing $C_{|\psi\rangle,\thinspace \hat{a}\hat{a}^{\dagger}}(t)$, it is helpful first to examine how $\hat{a}(t)$ depends on $t$. A single application of $\hat{U}_\tau$ on $\hat{a}$ gives
\begin{eqnarray}
\hat{a}(1)=\hat{U}^{\dagger}_{\tau}\hat{a}\hat{U}_{\tau}=e^{-2\pi i/R}\left[\hat{a}+\dfrac{iK}{\sqrt{2\hbar\omega}}\sin\hat{X}\right].
\end{eqnarray}
After $t$ recursive applications of $\hat{U}_{\tau}$, the Heisenberg evolution of $\hat{a}$ reads as
\begin{eqnarray}\label{bosonic-evolution1}
\hat{a}(t)e^{2\pi i t/R}=\hat{a}+\dfrac{iK}{\sqrt{2\hbar\omega}}\sum_{j=0}^{t-1}e^{2\pi ij/R}\sin\hat{X}(j),
\end{eqnarray}
where $\hat{X}(j)=\hat{U}^{\dagger j}_{\tau}\hat{X}\hat{U}^j_{\tau}$ and $\sin\hat{X}=[e^{i\hat{X}}-e^{-i\hat{X}}]/2i$. From Eq. (\ref{bosonic-evolution1}), we make the following immediate observations: (i) the total number of terms on the right-hand side scales linearly with $t$ and (ii) the operator $\sin\hat{X}(j)$ is always bounded, i.e., $\|\sin\hat{X}(j)|\psi\rangle\|\leq|\|\psi\rangle\|$ for any $|\psi\rangle$. As a result, the asymptotic growth of  $\hat{a}(t)$ can be at most linear. Long-time dynamics of several other quantities, such as mean energy growth, follow directly from this result. For instance, in an arbitrary Fock state $|n\rangle$, one can use Eq. (\ref{bosonic-evolution1}) to show that the mean energy is always bounded above by a quadratic function of $t$, i.e.,
\begin{equation}
\langle n|(a^\dagger a)(t)|n\rangle\leq n+\sqrt{\dfrac{2n}{\hbar\omega}}Kt+\dfrac{K^2t^2}{2\hbar\omega}.
\end{equation}

In a similar way, we use Eq. (\ref{bosonic-evolution1}) to learn the behavior of $C_{|\psi\rangle,\thinspace \hat{a}\hat{a}^{\dagger}}(t)$. To do so, we take
\begin{eqnarray}\label{com}
\left[\hat{a}(t), \hat{a}^{\dagger}\right]e^{2\pi it/R}=1+\dfrac{iK}{\sqrt{2\hbar\omega}}\sum_{j=0}^{t-1}e^{2\pi ij/R}\left[\sin\hat{X}(j), \hat{a}^{\dagger}\right].
\end{eqnarray}
For $K=0$, the system is just an integrable harmonic oscillator and the OTOC remains a constant in any given state $|\psi\rangle$ for all $t\geq 0$ --- $C_{|\psi\rangle,\thinspace \hat{a}\hat{a}^{\dagger}}(t)=1$. This means that the initial operator $\hat{a}$ remains fully regular, i.e., the operator $\hat{a}(t)$ retains the diagonality in the coherent state basis. For $K\neq 0$, as the system evolves, the operator will start to scramble into the operator Hilbert space via the mixing of eigenstates, which is hinted at by the positive growth of the OTOC. For example, after one time step, we can calculate explicitly that
\begin{equation}\label{first_step}
C_{|\psi\rangle,\thinspace \hat{a}\hat{a}^{\dagger}}(t=1)=1+\dfrac{K^2}{4\omega^2}\langle\psi|\cos^2\hat{X}|\psi\rangle\geq 1\text{ (for all }|\psi\rangle).
\end{equation}
For $t>1$, in general, a closed form expression for $C_{|\psi\rangle,\thinspace \hat{a}\hat{a}^{\dagger}}(t)$ is out of reach. Hence, we resort to numerical methods to probe the OTOCs. To be precise, we numerically compute the OTOCs by considering a weak ($K\lesssim 1$) and a moderately strong ($K\sim O(1)$) kicking strength under both resonance ($R=4$) and non-resonance ($R=3.9$) conditions. We choose two initial quantum states: the vacuum state $|0\rangle$ and a coherent state $|\alpha\rangle$. In what follows, we shall first examine the early-time behavior of the OTOCs, then proceed to analyse the long-time dynamics.
\begin{figure*}
\includegraphics[width=\textwidth,height=6.5cm]{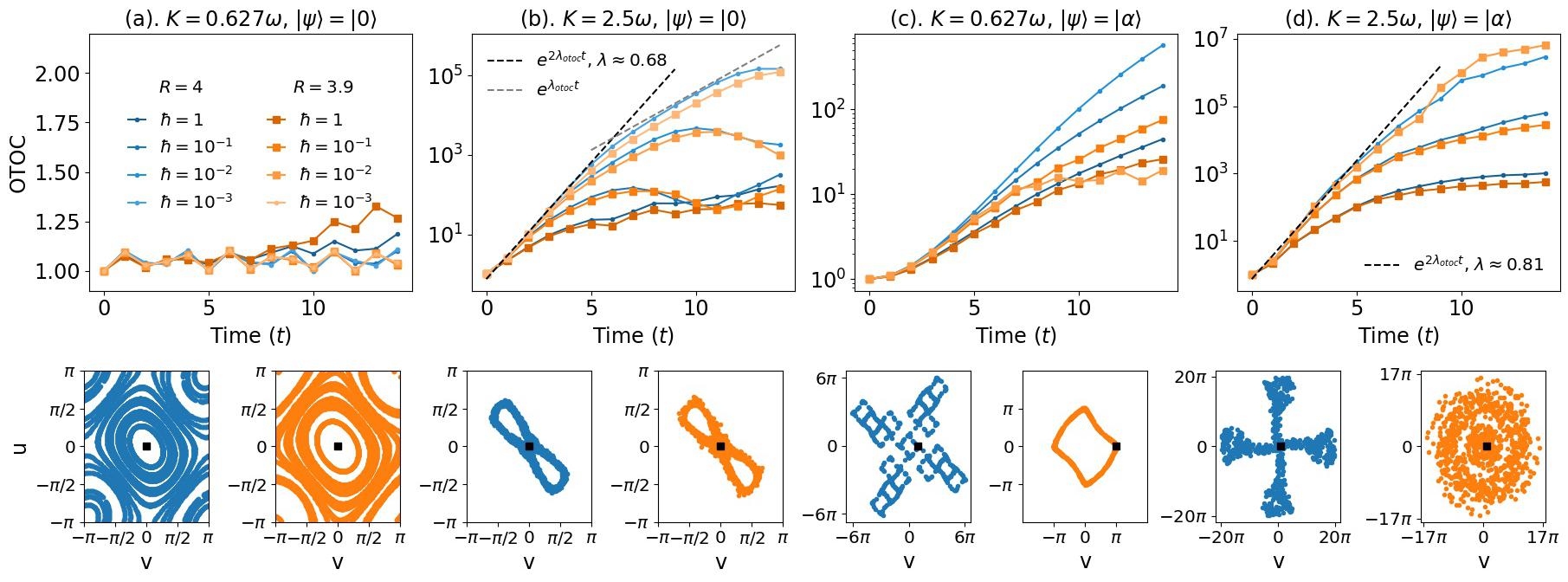}
\caption{\label{fig:protoc} The figure illustrates the early time OTOC behavior near $R=4$ in two different perturbative regimes of the quantum KHO, considering two different initial states, namely, the vacuum state $|0\rangle$ and a coherent state $|\alpha\rangle$, where $\Re(\alpha)=\sqrt{\omega/2\hbar}\langle\hat{X}\rangle$ and $\Im(\alpha)=\langle\hat{P}\rangle/\sqrt{2\omega\hbar}$, centered at an unstable period four point given by $(\langle\hat{P}\rangle, \langle\hat{X}\rangle)=(0, \pi)$. The kicking period is fixed at $\tau=1$, which automatically sets $\omega=2\pi/R$. While the top panels demonstrate the OTOC calculations, the bottom panels show the classical phase space dynamics in the vicinity of the corresponding initial conditions. The square-shaped black dot in the bottom panels represents the corresponding classical initial condition. Also, note the color coding --- we use blue to denote the resonant case ($R=4$) and orange for the non-resonant case ($R=3.9$). The darker colors denote the deep quantum regime, and the lighter colors indicate the semiclassical domain. (a)-(b) The system is initialized in the vacuum state $|0\rangle$. In (a), the kicking strength is fixed at $K=0.627\omega$, corresponding to the weak perturbative regime. Moderately strong perturbation is considered in (b). The plot shows early-time dynamics on the semi-log scale. Black and grey colored dashed lines are plotted here to illustrate the two-step early-time exponential growth. Furthermore, we find no visible differences between the cases of $R=4$ and $R=3.9$. (c)-(d) Here, we repeat the same calculations as before by replacing the vacuum state with the coherent state [see the main text for details].}
\end{figure*}

\subsection{Early-time dynamics}
\label{resonanceOTOC}
Classically, the vacuum state corresponds to a fixed point of the dynamical map given in Eq. (\ref{dynmap}). In contrast, the coherent state is chosen to be centered at a point on the classical stochastic web associated with the resonant case $R=4$, satisfying the equation $v=\pm(u+\pi)+2l\pi$, $l\in\mathbb{Z}$. In the phase space, these two initial conditions give rise to different dynamics altogether. Thus, it is essential to see if the differences are also reflected in the behavior of the OTOCs over short periods.
\subsubsection{Vacuum state (fixed point) OTOC}
We first discuss the resonant scenario and then contrast it with the non-resonant one. Due to the correspondence principle, the early time growth of the vacuum state OTOCs will have a close correspondence with the fixed point behavior of $(u, v)=(0, 0)$ in the classical phase space. To illustrate, we consider the Jacobian matrix of the classical map evaluated at $(0, 0)$. 
\begin{equation}\label{jacob}
J_{R=4}=
\left.\begin{bmatrix}
0 & 1 \\
-1 & -\dfrac{K\cos{v_n}}{\omega}
\end{bmatrix} \right|_{u_n=0, v_n=0}=
\begin{bmatrix}
0 & 1 \\
-1 & -\dfrac{K}{\omega}
\end{bmatrix},
\end{equation}
whose eigenvalues are
\begin{math}
\gamma_\pm =(-K\pm\sqrt{K^2-4\omega^2})/2\omega .
\end{math}
In this work, we always assume that $K$ is positive. Therefore, for $K<2\omega$, the eigenvalues are a pair of complex conjugates with unit modulus, i.e.,  $|\gamma_{\pm}|=1$, which implies that the fixed point is stable. The phase space trajectories in this region will wrap around $(0, 0)$ in closed elliptically-shaped orbits. Consequently, the vacuum state OTOCs are expected to remain stagnant until the Ehrenfest time due to the vanishing Lyapunov exponent. The map then undergoes a bifurcation at $K=2\omega$ with the emergence of two other stable fixed points. For $K>2\omega$, the point $(0, 0)$ becomes a saddle --- in this case, the trajectories with the initial conditions located slightly off $(0, 0)$ will diverge exponentially from one another with the saddle point exponent $\lambda_s$ given by $\sim\text{max}\{\log(|\gamma_+|), \log(|\gamma_-|)\}$. Accordingly, the OTOCs in the vacuum state $|0\rangle$ or any coherent state $|\alpha\rangle$ that lives close by are expected to display short-time exponential growth whenever $K>2\omega$. 

Figure \ref{fig:protoc}a and \ref{fig:protoc}b illustrate the vacuum state OTOCs for two different kicking strengths, namely, $K=0.627\omega$ and $K=2.5\omega$ in the vicinity of $R=4$. All the blue curves (with point markers) correspond to $R=4$, and the orange-colored curves (with square markers) represent the case of $R=3.9$. First, when $R=4$ and $K$ is small, the function $C_{|0\rangle,\thinspace \hat{a}\hat{a}^{\dagger}}(t)$ shows an initial stagnant behavior, which is demonstrated in Fig. \ref{fig:protoc}a. The lack of growth at initial times is reminiscent of the classical elliptic stability of the point $(0, 0)$. On the other hand, for $K=2.5\omega$, we observe the initial growth to be quadratic in the deep quantum regime ($\hbar=1$) [see Fig. \ref{fig:protoc}b]. This is because, in the deep quantum regime, the Ehrenfest time $\tau_{\text{EF}}\sim\ln(1/\hbar_{\text{eff}})/2\lambda_s$ \cite{schubert2012wave} remains very small, which makes it hard to observe the exponential growth. However, one can witness genuine exponential growth by slowly moving towards the semiclassical regime. This can be done by tuning the Planck constant $\hbar$. While for $K=0.627\omega$, the semiclassical OTOCs under the resonance remain stagnant for much longer as they should be, the case of $K=2.5\omega$ shows a clear exponential scaling. In the latter case, when $\hbar=10^{-3}$, the quantum Lyapunov exponent extracted from the OTOC through an exponential fitting of the first six data points is $\lambda_{\text{otoc}}\approx 0.68$. This value aligns well with the classical saddle point exponent of the origin, $\lambda_s=\ln(2)\approx 0.693$. In Ref. \cite{steinhuber2023dynamical}, it has been observed that in locally hyperbolic systems, the OTOCs exhibit a two-step early-time exponential growth. Due to the hyperbolicity of $(0, 0)$, a similar behavior is expected in the vacuum state OTOC for $K > 2\omega$. We indeed observe in Fig. \ref{fig:protoc}b that the initial growth of $\sim e^{2\lambda_{\text{otoc}}t}$ is followed by a subsequent regime scaling as $\sim e^{\lambda_{\text{otoc}}t}$.

To strengthen the correspondence between the classical and the quantum exponents, we analytically extract the quantum exponent from the vacuum state OTOC in the semiclassical limit ($\hbar\rightarrow 0$). Let $l \in \mathbb{R}^+$ and $l>1$, then reducing $\hbar$ by the factor of $l$ is equivalent to increasing both $K$ and $\omega$ in the second term of the Floquet operator by the same factor $l$ while keeping $\hbar$ fixed. This, however, does not affect the first term $\exp\{-i2\pi/R \hat{a}^{\dagger}\hat{a}\}$ \footnote[1]{We thank the anonymous referee(s) for pointing this out}. Since $K$ is now increased, one plausible effect would be enhanced scrambling. Then again, for small $t$, with the initial state $|0\rangle$, it holds that $\cos(\hat{X}/\sqrt{l})\approx 1-\hat{X}^2/2l$ for $l\gg 1$. Therefore, we have $e^{-iKl\cos(\hat{X}/\sqrt{l})}\approx e^{-iKl}e^{iK\hat{X}^2/2}$. The phase term $e^{-iKc}$ can be ignored here. At $R=4$, under the modified Floquet evolution, the bosonic operators evolve according to the following transformation:
\begin{eqnarray}
\begin{bmatrix} \hat{a}(t) \\ \hat{a}^{\dagger}(t) \end{bmatrix}
 =
\begin{pmatrix}
   y-i & y \\
   y & y+i
\end{pmatrix}^{t}
\begin{bmatrix} \hat{a} \\ \hat{a}^{\dagger} \end{bmatrix}, 
\end{eqnarray}
where $y=K/2\omega$. 
The eigenvalues of the above linear transformation are given by $\lambda_{\pm}=(-K\pm\sqrt{K^2-4\omega^2})/2\omega$, which also turn out to be the eigenvalues of the Jacobian in Eq. (\ref{jacob}). Hence the semiclassical vacuum state OTOC ($C_{|0\rangle ,\hat{a}\hat{a}^{\dagger}}(t)$) grows exponentially in time with $\text{max}\{\log(|\lambda_+|), \log(|\lambda_-|)\}$ being the rate of growth. For $K>2\omega$, the growth remains exponential. For $K=2\omega$, the classical bifurcation point, $|\lambda_{\pm}|=1$, leading to the stagnant behavior. When $K$ is below $2\omega$, oscillatory behavior is expected. This establishes a strong classical-quantum correspondence in the limit of $\hbar\rightarrow 0$. At this point, the time scale for which the above approximation remains valid is unclear. Nevertheless, given the nature of classical fixed point behavior, we anticipate its validity within the Ehrenfest regime.

Interestingly, for $R=3.9$, we observe that the short-time growth of the OTOCs almost coincides with that of the resonant counterpart, regardless of the strength of kicking. This can be intuitively understood by examining the fixed point nature of ($0, 0$) under the non-resonance condition. The general condition for the bifurcation of $(0, 0)$ is given by
\begin{equation}\label{bif}
\dfrac{K}{\omega}=\dfrac{2\cos\omega\tau\pm 2}{\sin\omega\tau}, \text{ where }\omega\tau=\dfrac{2\pi}{R}.
\end{equation}
Accordingly, for $R=3.9$, the bifurcation point is located at $K= 1.921\omega$, slightly off that of $R=4$. Therefore, for $K<1.921\omega$, the point $(0, 0)$ remains stable. Also, when the perturbation is more than what is required for the bifurcation, the origin becomes a saddle with the emergence of two other stable fixed points. Thus, the bifurcation mechanism resembles that of the resonance, with the only difference being a slight change in the bifurcation location along the parameter axis of $K$. Consequently, at any given perturbation strength, the phase trajectories in the vicinity of the origin do not acquire any major changes as $R$ is moved from $4$ to $3.9$. The phase space plots presented in Fig. \ref{fig:protoc}a support this conclusion. Therefore, we expect the short-time growth of the vacuum state OTOCs for $R=3.9$ and $R=4$ to be qualitatively identical, irrespective of other system parameters, which is confirmed by the numerical results. Similar to the case of $R=4$, the two-step exponential growth is also observed for $R=3.9$. This behavior is evident in Fig. \ref{fig:protoc}b, where the blue and orange curves exhibit nearly identical growth rates.

\subsubsection{Coherent state (on the web) OTOCs}
As for the coherent state OTOCs, at $R=4$, the short-time dynamics rely mainly on the nature of the classical stochastic web. Recall that for $K<2\omega$, the origin ($0, 0$) is the only classical fixed point. There exist, however, an infinite number of period four points, each located at $(u, v)=(p\pi, q\pi)$ with $p, q\in\mathbb{Z}$ for any arbitrary small $K$ \cite{kells2005quantum}. The period four orbits these points generate are stable if $p+q$ is even and unstable otherwise. Thus, the presence of alternate stable and unstable manifolds causes the movement of any trajectory set out from the neighborhood of an unstable orbit to be highly complex. Such a complex motion persists even in the limit of small $k$, where the Lyapunov exponent approaches zero --- for example, see the phase space trajectories in Fig. \ref{fig:poincare}d. We are interested in seeing if such behavior is also reflected in coherent state OTOCs. Here, for the OTOC calculations, we focus on a specific coherent state with the mean coordinates located at $(\langle\hat{P}\rangle, \langle\hat{X}\rangle)=(0, \pi)$, which corresponds to an unstable period four point. 

The results are plotted in Fig. \ref{fig:protoc}c and \ref{fig:protoc}d for the same perturbation strengths as before. We first discuss the case of resonance. When $K=0.627\omega$, the OTOC in the deep quantum regime exhibits an approximate quadratic growth. In this case, the corresponding classical phase space trajectories diverge at a rate given by $\lambda_{\text{cl}}\approx 0.06$, the maximum Lyapunov exponent associated with the unstable period four point $(0, \pi)$. Due to the smallness of the exponent, it's difficult to observe the exponential growth even when $\hbar$ is small. On the other hand, when $K=2.5\omega$, the corresponding classical exponent is given by $\lambda_{\text{cl}}\approx 0.9$. In this case, the OTOC in the deep quantum regime ($\hbar =1$) still displays an initial algebraic growth. However, upon reducing $\hbar$, a clean exponential growth can be observed. For $\hbar=10^{-2}$, the rate of growth of the OTOC is extracted to be $\lambda_{\text{otoc}}\approx 0.81$, which agrees closely with the classical exponent, highlighting a good quantum-classical correspondence. The corresponding plots are shown in Fig. \ref{fig:protoc}d. We also notice that the coherent state OTOCs for smaller $\hbar$ overshoot those for larger $\hbar$ in both panels \ref{fig:protoc}c and \ref{fig:protoc}d. Moreover, since the stochastic webs fill a significant portion of the phase space at resonances, it is expected that nearly all coherent states overlapping with these webs will exhibit similar behavior as shown in Fig. \ref{fig:protoc}c and \ref{fig:protoc}d.

\begin{figure*}
\includegraphics[width=\textwidth,height=4.cm]{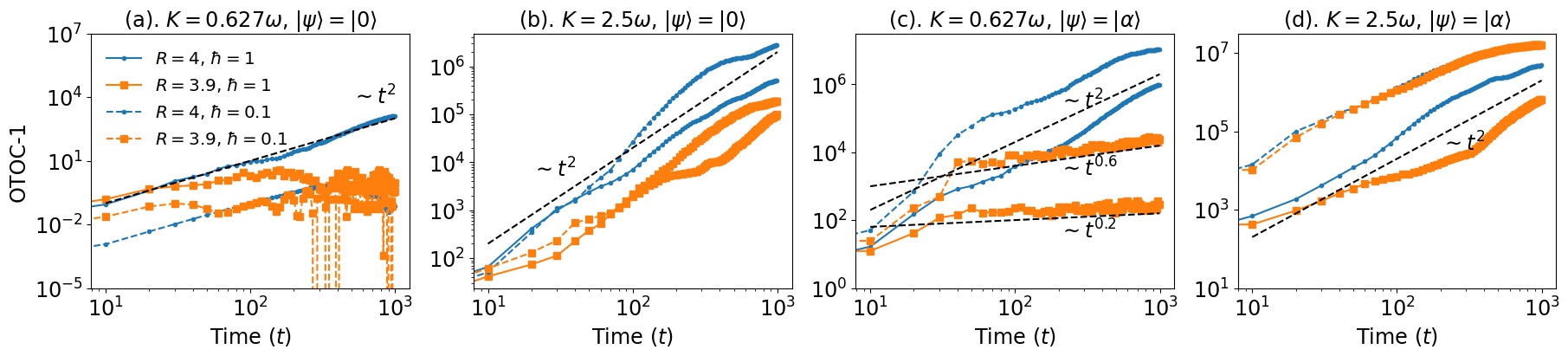}
\caption{\label{fig:protoc_ext} The figure illustrates the OTOC growth over a period of $10^3$ time steps for both the initial states $|0\rangle$ and $|\alpha\rangle$. We consider the interval between two successive data points $10\tau$ to avoid computational overhead due to the unbounded nature of the system. The parameters across the panels are fixed to be the same as those in Fig. \ref{fig:protoc}. The black-colored dashed lines following various scalings are drawn to contrast the growth under the resonance and non-resonance conditions. In the strong perturbative regime ($K=2.5\omega$), the quadratic growth remains the dominant behavior for all the parameters as shown in \ref{fig:protoc_ext}b and \ref{fig:protoc_ext}d. The differences between the resonant and non-resonant cases become apparent when $K$ is small [see the main text for details. Also see Fig. \ref{fig:protoc} for the corresponding short-time growth]. Note that as the time interval is large, the oscillations may not be fully visible in Fig. \ref{fig:protoc}a.}
\end{figure*}

To contrast the above results with the non-resonant case, we repeat the OTOC calculations by setting $R=3.9$ for the same coherent state $(\langle \hat{P}\rangle, \langle\hat{X}\rangle)=(0, \pi)$. First, under small perturbations, the classical phase space remains primarily regular. Hence, the initial time growth of OTOCs is expected not to contain any exponential scaling. We confirm this from Fig. \ref{fig:protoc}c, where the initial growth (as shown by orange curves with square markers) for all the cases of $\hbar$ is algebraic and shows no exponential growth. At initial times, the orange curves nearly coincide with the blue curves. However, a clear departure from one another can be observed as time progresses. The algebraic growth can be reasoned with the following argument: Although the system is in non-resonance, the closeness of $R$ to the resonance condition suggests a competition between the diffusive and regular dynamics. The diffusive dynamics usually dominate the early-time behavior. Hence, we see a coincidence between the initial growths of resonant and non-resonant cases. On the other hand, when $K$ is sufficiently large, the classical phase space trajectories display chaotic diffusion. One might expect exponential scaling over a short time in this case. The numerics indicate that the OTOCs in the deep quantum regime still display algebraic growth. Nevertheless, the exponential growth appears in the semiclassical regime [see Fig. \ref{fig:protoc}d], where the orange curves show identical growth as the blue curves. Recall that reducing $\hbar$ leads to a competition between the enhanced scrambling and the oscillatory behavior due to an effective increase in $K$ and $\omega$. Here, for $\hbar=0.1$, the OTOC grows at a faster rate than the case of $\hbar=1$ for both values of $K$. There is, however, no reason to expect similar behavior as $\hbar$ further varies. For instance, in Fig. \ref{fig:protoc}c, the growth for $\hbar=0.01$, as shown by the lighter orange curve, seems to be suppressed after some initial time. Intuitively, the correspondence principle implies oscillatory behavior in the OTOC as $\hbar\rightarrow 0$ as the phase space is stable under non-resonances. Hence, for very small $\hbar$, the enhanced scrambling effect is suppressed, and the oscillatory behavior takes over. When the perturbation is large, the semiclassical OTOCs outperform those in the deep quantum regime [see Fig. \ref{fig:protoc}d].

\subsection{Long-time dynamics}
\label{non-resonanceOTOC}
Although the OTOCs do not appear to distinguish the non-resonances from the resonances over a short time, the long-time dynamics produce significant differences between the two. These differences are more pronounced when the kicking strength is small ($K\lesssim 1$). Let us first examine the resonant cases to see how long-time dynamics stand out. Recall from Eq. (\ref{com}) that the total number of terms in the equation grows linearly with time. At each time step, the right-hand side grows by one more term, which is given by $[\sin\hat{X}(j), \hat{a}^{\dagger}]e^{2\pi ij/R}$. The resonances will then lead to a \textit{coherent} addition of all the terms, thus rendering the asymptotic growth of the OTOC a quadratic function of time, for a typical initial state --- $C(t)\sim K^2t^2$. When $K$ is too small ($K\ll 1$), the operator $\sin\hat{X}(j)$ grows only by a negligible amount. Then, the OTOC can be explicitly shown to exhibit quadratic growth by ignoring the terms of order $O(K^3)$ and higher in the time-evolved operator $\hat{a}(t)$. Refer to Appendix \ref{appendix:b} for more details, where we give an explicit derivation for the same. However, since $\sin\hat{X}(j)$ is bounded, the quadratic growth is expected to persist even when $K$ is large.

To verify numerically, we refer to Fig. \ref{fig:protoc_ext}, where the long-time dynamics have been illustrated for $\sim 10^3$ time steps. Due to the unbounded nature of the system, a sufficiently large Hilbert space is needed to perform these numerical simulations. Here, we analyze the resonant cases in the figure. The figure demonstrates that in the deep quantum regime ($\hbar=1$), for both the initial states $|0\rangle$ and $|\alpha\rangle$, the long-time dynamics always scale quadratically as long as the resonance condition is satisfied regardless of the strength of perturbation. While the same result seems to hold in the semiclassical limit for the coherent state, the vacuum state OTOC shows anomalous oscillatory behavior as shown in Fig. \ref{fig:protoc_ext}a. This behavior is not surprising as it is expected due to the classical-quantum correspondence. The classical stability of $(0, 0)$ implies a longer $t_{\text{EF}}$ for the vacuum state OTOC in the semiclassical limit compared to other cases in the figure. The numerics suggest that this behavior persists beyond $10^3$ time steps. The quadratic scaling may emerge as the OTOC picks up non-trivial $\hbar$ effects. Moreover, for large $K$, the quadratic growth persists for both $\hbar=1$ and $\hbar=0.1$ as shown in Fig. \ref{fig:protoc_ext}b and \ref{fig:protoc_ext}d. It is worth noting that the degree of scrambling, as quantified by the magnitude of the OTOC, always depends on the specific choice of the state vector despite both states exhibiting long-time quadratic growth. For example, the coherent states with the mean coordinates located on the stochastic web delocalize quickly in the phase space when acted upon by the Floquet unitary $\hat{U}_{\tau}$. In these states, the initial operators are relatively more prone to get scrambled compared to the vacuum state $|0\rangle$.

Contrary to the resonant cases, the non-resonant $R$ comprises a dense set of rational numbers (excluding integers) with measure zero and irrational numbers with measure one. When $R$ is irrational, the equidistribution property implies that in the long-time limit, the phases $\{e^{2\pi ij/R}\}_{j=0}^{t}$ will tend to behave as though they were drawn uniformly at random from the unit circle in the complex plane. This leads to \textit{incoherent} summations, causing various terms in Eq. (\ref{com}) to interfere destructively. Therefore, we expect that the growth of the OTOC will be suppressed. As we shall show in the next section, the irrationality of $R$, on average, induces linear growth in the OTOC [see also Appendix \ref{appendix:b} for more details]. Let us also point out that the OTOC growth in a typical pure state may generally vary from the linear behavior. Moreover, the rational $R$ values also induce subdued growth, though not as much as the irrationals. The results demonstrated in Fig. \ref{fig:protoc_ext} indicate that the growth is suppressed for $R=3.9$ in all the cases considered (shown in the orange-colored curves). When $K$ is small, while the vacuum state OTOCs oscillate as shown in Fig. \ref{fig:protoc_ext}a, the coherent state OTOCs follow power laws given by $\sim t^{0.2}$ and $\sim t^{0.6}$ respectively for $\hbar=1$ and $\hbar=0.1$ [see Fig. \ref{fig:protoc_ext}c]. As $K$ becomes large, the OTOCs at non-resonances will also grow quadratically as the system becomes fully chaotic.    
\section{Analytically Tractable cases}
\label{analtytical}
In the preceding section, our analysis primarily relied on numerical investigations to compare the operator growth in both the resonant and the non-resonant scenarios. Given that the kicking potential is highly non-linear, the exact solutions of the quantum KHO are generally intractable. There is, however, a narrow window of opportunity involving a few special cases, such as the symmetries and \textit{quantum resonances} that allow for the analytical treatment of the OTOCs. In this section, we exploit the translation invariance of the quantum  KHO by considering a few selective resonant cases to obtain the OTOCs as explicit functions of $t$. In particular, we consider the trivial cases, $R=1$ and $2$, followed by a more intricate case of $R=4$. Additionally, we provide an analytical approximation for the OTOC averaged over the space of pure states, which we shall refer to as the average state-OTOC, by considering small $K$ ($K\ll 1$) and an irrational $R$. 
\subsection{Case-1: $R=1$ and $2$}
Here, we focus on the commutator function $C_{|\psi\rangle,\thinspace \hat{a}\hat{a}^{\dagger}}(t)$ for the case of $R=2$. We do not consider $R=1$ separately since the resulting expressions for the OTOCs are the same in both cases. In the present case, stochastic webs do not constitute the classical phase space due to the exact solvability of the classical map, which is given by
\begin{eqnarray}\label{cla-R2}
v_{n} &=&(-1)^nv_0\nonumber\\ 
u_{n} &=&(-1)^n\left[u_0+n\varepsilon\sin v_0\right].
\end{eqnarray} 
It is worth noting that the time-evolved operators $\hat{X}(t)$ and $\hat{P}(t)$ admit the same functional form as the classical variables $v_n$ and $u_n$, indicating a persistent classical-quantum correspondence. 

When $R=2$, the phase operator $e^{-(2\pi i/R) \hat{a}^{\dagger}\hat{a}}$ possesses alternating $+1$ and $-1$ on its diagonal. This enables us to express $\hat{a}(t)$ explicitly as 
\begin{equation}
(-1)^t\hat{a}(t)=\hat{a}+\dfrac{iKt}{\sqrt{2\hbar\omega}}\sin\hat{X},
\end{equation}
which shows a clear linear dependence on $t$ accompanied by an oscillatory behavior originating from the term $(-1)^t$. By noting that $[\sin\hat{X}, \hat{a}^{\dagger}]=\sqrt{\hbar/2\omega}\cos\hat{X}$, we finally obtain
\begin{eqnarray}\label{otocatR2}
C_{|\psi\rangle,\thinspace \hat{a}\hat{a}^{\dagger}}(t)=1+\dfrac{K^2t^2}{4\omega^2}\langle\psi|\cos^2\hat{X}|\psi\rangle,
\end{eqnarray}
which is a quadratic function of $t$. The oscillating term $(-1)^t$ disappears as the OTOCs involve absolute squares of the commutators. Moreover, the commutator function always retains quadratic growth except when $\langle\psi|\cos^2\hat{X}|\psi\rangle=0$. As an example, we here take $|\psi\rangle=|0\rangle$, then $\langle 0|\cos^2\hat{X}|0\rangle=e^{-\hbar/2\omega}\cosh(\hbar/2\omega)$. We can therefore write
\begin{equation}
C_{|0\rangle,\thinspace \hat{a}\hat{a}^{\dagger}}(t)=1+\dfrac{K^2t^2}{4\omega^2}e^{-\hbar/2\omega}\cosh\left(\dfrac{\hbar}{2\omega}\right).
\end{equation}

To obtain the OTOC as a state-independent quantity, one can average $C_{|\psi\rangle,\thinspace \hat{a}\hat{a}^{\dagger}}(t)$ over the space of pure states or a suitable set of vectors forming a continuous variable $1$-design. In continuous variable systems, the Fock state and the coherent state bases are known to form $1$-designs \cite{blume2014curious, iosue2022continuous}. Note that in finite-dimensional systems, the OTOCs are often evaluated in the maximally mixed states, equivalent to averaging over the Haar random pure states. On the contrary, the notion of maximally mixed states in continuous variable systems is not well-defined due to the diverging traces induced by the infinite-dimensional Hilbert spaces. Despite this limitation, the averaging procedure can still provide insights into the nature of scrambling in a typical pure state.

For $R=2$, one can readily evaluate the average state-OTOC as follows:
\begin{eqnarray}
\overline{C_{|\psi\rangle,\thinspace \hat{a}\hat{a}^{\dagger}}}(t)=1+\dfrac{K^2t^2}{4\omega^2}\overline{\cos^2\hat{X}}.
\end{eqnarray}
This expression can be simplified by recalling that $\cos\hat{X}$ is related to the displacement operators.
\begin{equation}
\cos\hat{X}=\dfrac{1}{2}\left[D\left(i\sqrt{\dfrac{\hbar}{2\omega}}\right)+D\left(-i\sqrt{\dfrac{\hbar}{2\omega}}\right)\right].
\end{equation}
From Eq. (\ref{dispavg}) and (\ref{dispavgfinal}), it then follows that
\begin{eqnarray}
\overline{C_{|\psi\rangle,\thinspace \hat{a}\hat{a}^{\dagger}}}(t)&=&1+\dfrac{K^2t^2}{16\omega^2}\left[\overline{D\left(i\sqrt{\dfrac{2\hbar}{\omega}}\right)+D\left(-i\sqrt{\dfrac{2\hbar}{\omega}}\right)+2D(0)}\right]\nonumber\\
&=&1+\dfrac{K^2t^2}{8\omega^2}.
\end{eqnarray}
In the second equality, we used the result from Appendix \ref{appendix:a} that $\overline{D(\beta)}=\delta_{\Re(\beta), 0}\delta_{\Im(\beta)}$ for any $\beta\in\mathbb{C}$. We also assumed a finite non-zero value for $\hbar$. The above expression for $\overline{C_{|\psi\rangle,\thinspace \hat{a}\hat{a}^{\dagger}}}(t)$ is exact and valid for all kicking strengths.

\subsection{Case-2: $R=4$}
Here, we set $R=4$ and derive the analytical expression for the OTOC owing to certain restrictions on $\omega$. Our analysis can also be extended to the other cases when $R=3$ and $6$. Recall from Ref. \cite{borgonovi1995translational} and Section \ref{section-2} that under the translational invariance, $R$-th powers of the unitary evolution, $U_{\tau}^R$, commute with either one-parameter or two-parameter groups of translations depending on the values $\omega$ assumes. Since $\hat{a}(t)$ contains the terms $\sin\hat{X}(j)$ for $j=0, ..., t-1$, we are interested in finding a suitable $\omega$ that ensures the commutation relation $[\sin{\hat{X}}, \hat{U}^4]=0$. In order to proceed, it is useful to write $U^4_{\tau}$ in a compact form as follows:
\begin{equation}
\hat{U}^4_{\tau}=\left(e^{-i(K/\hbar)\cos\left(\hat{P}/\omega\right)}e^{-i(K/\hbar)\cos{\hat{X}}}\right)^2,
\end{equation}
where the Floquet operator within the parentheses can be identified with the kicked Haper model \cite{kells2005eigensolutions}. Since $\sin\hat{X}$ always commutes with $\cos\hat{X}$, we now only require to find $\omega$, that allows for the commutation between $\sin\hat{X}$ and $\cos(\hat{P}/\omega)$, i.e, $[\sin\hat{X}, \cos(\hat{P}/\omega)]=0$.

One can verify that the commutator $[\sin\hat{X}, \cos(\hat{P}/\omega)]$ indeed vanishes whenever $\omega=\hbar/2k\pi$ with $k\in\mathbb{Z}^+$. To see this, we consider the following:
\begin{eqnarray}
\left[e^{i\hat{X}}, e^{\pm i\hat{P}/\omega}\right]&=&e^{i\hat{X}}e^{\pm i\hat{P}/\omega}-e^{\pm i\hat{P}/\omega}e^{i\hat{X}} \nonumber\\
&=&e^{i\hat{X}\pm (i\hat{P}/\omega)\mp (i\hbar/2\omega)}-e^{i\hat{X}\pm (i\hat{P}/\omega)\pm (i\hbar/2\omega)}\nonumber\\
&=&(-1)^k\left[e^{i\hat{X}\pm 2ki\pi\hat{P}}-e^{i\hat{X}\pm 2ki\pi\hat{P}}\right]\hspace{0.3cm}\text{for }\omega=\dfrac{\hbar}{2k\pi}\nonumber\\
&=&0.
\end{eqnarray}
The second equality follows from the BCH formula. In the third equality, we have made the substitution that $e^{\pm ik\pi}=(-1)^k$. Likewise, $e^{-i\hat{X}}$ can also be shown to commute with $\cos(\hat{P}/\omega)$. Alternatively, one can also show that the operator $e^{\pm i\hat{X}}$ belongs to a one-parameter group of translations whenever the above condition on $\omega$ is satisfied \cite{borgonovi1995translational}. As a result, we have $\hat{U}^{\dagger 4}_{\tau}\sin\hat{X}\hat{U}^{4}_{\tau}=\sin\hat{X}$. Therefore, for some $j\in \{0, 1, 2, 3\}$, the translational invariance implies $\sin\hat{X}(j)=\sin\hat{X}(t+j)$, for $t$ being an integer multiple of $4$. Furthermore, due to $\hbar$ dependence, this is also called the quantum resonance condition \cite{billam2009quantum}. In this case, the operator $\hat{a}(t)$ at a total time $t=4s$, where $s$ is a non-negative integer, can be expressed as an explicit function of $t$:
\begin{eqnarray}\label{R4bos}
\hat{a}(t)=\hat{a}+\dfrac{iKt}{4\sqrt{2\hbar\omega}}\sum_{j=0}^3 e^{ij\pi/2}\sin\hat{X}(j).
\end{eqnarray}
This allows us to write the OTOC as follows: 
\begin{align}\label{otocat4res}
C_{\hat{a}\hat{a}^{\dagger}}&(t)=1+\dfrac{Kt}{4\sqrt{2\hbar\omega}}\left[\sum_{j=0}^3i^{j+1}\left[\sin\hat{X}(j), \hat{a}^{\dagger}\right]+\textbf{h.c.}\right]\nonumber\\
&+\dfrac{K^2t^2}{32\hbar\omega}\left[\sum_{j, j'=0}^{3}i^{j'-j}\left[\sin\hat{X}(j), \hat{a}^{\dagger}\right]^{\dagger}\left[\sin\hat{X}(j'), \hat{a}^{\dagger}\right]\right],
\end{align}
where $\textbf{h.c.}$ denotes the Hermitian conjugate of the operator $i^{j+1}[\sin\hat{X}(j), \hat{a}^{\dagger}]$.
\begin{figure}
\includegraphics[scale=0.35]{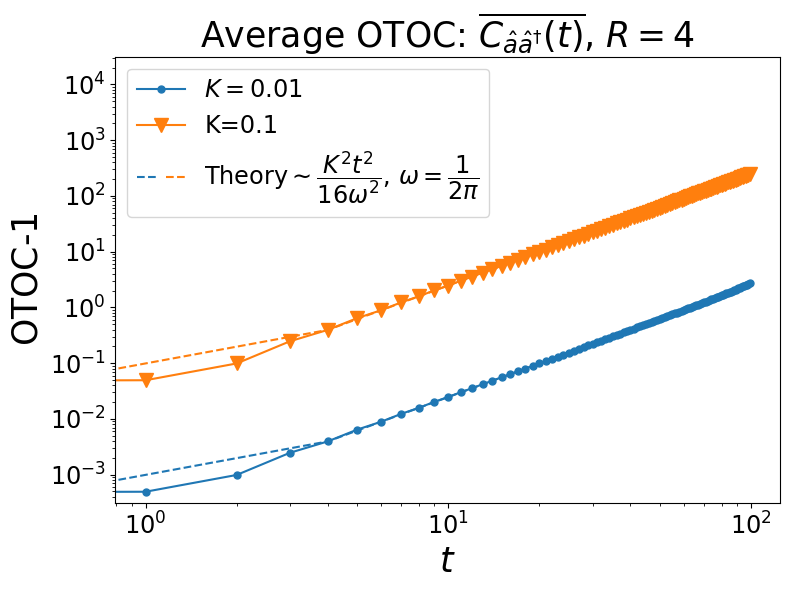}
\caption{\label{fig:otoct} The averaged commutator function $\overline{C_{\hat{a}\hat{a}^{\dagger}}(t)}$ for the translation invariant case at $R=4$. The figure illustrates the growth for two values of $K$, namely, $K=0.01$ and $0.1$. We have taken $\hbar=1$. While the solid lines with the markers denote numerical results, the dashed lines correspond to the theory as given in Eq. (\ref{otocat4}). The solid and dashed lines almost coincide for all $t>0$. For the numerical results, we perform the average over two thousand number states.}
\end{figure}
Although Eq. (\ref{otocat4res}) implies that the growth of the OTOC is a quadratic polynomial of time, the presence of operators such as $\{\sin\hat{X}(j)\}_{j=1}^3$, which are not amenable to the exact treatment, makes the OTOC only quasi-exact solvable. Nevertheless, in the limit of weak perturbations, we can still obtain a good approximate analytical solution up to the leading orders in $K$. In particular, when $K$ is small ($K\ll\ 1$) and at sufficiently long times, the terms of the order $O(Kt)$ and $O(K^2t^2)$ dominate over the remaining others, such as $O(K^mt)$ with $m\geq 2$ and $O(K^nt^2)$ with $n\geq 3$. Therefore, we only require approximating the terms $\{\sin\hat{X}(j)\}$ to the zeroth order in $K$ --- $\sin\hat{X}(1)\approx i\sin (\hat{P}/\omega)\approx -\sin\hat{X}(3)$ and $\sin\hat{X}(2)\approx -\sin\hat{X}$. Thus, ignoring all the other insignificant contributions, we finally obtain
\begin{equation}\label{R4comf}
C_{\hat{a}\hat{a}^{\dagger}}(t)\approx 1+\dfrac{K^2t^2}{16\omega^2}\left[\cos\hat{X}+\cos\left(\dfrac{\hat{P}}{\omega}\right)\right]^2.
\end{equation}
Eq. (\ref{R4comf}) can be evaluated in any arbitrary initial state. We calculate the average state-OTOC to better understand its behavior in a typical quantum state. 
\begin{equation}\label{otocat4}
\overline{C_{|\psi\rangle,\thinspace \hat{a}\hat{a}^{\dagger}}(t)}= 1+ct^2\hspace{0.2cm}\text{for }t=4s, \hspace{0.1cm}s\in\mathbb{Z}^+\cup\{0\}, 
\end{equation}
where $c\approx K^2/16\omega^2$ for $K\ll 1$.

Figure \ref{fig:otoct} contrasts Eq. (\ref{otocat4}) with the numerically computed average state-OTOC for two different kicking strengths. The quantum resonance condition is invoked by fixing the frequency at $\omega=\hbar/2\pi$, with $\hbar=1$. We find an excellent agreement between the numerical results and Eq. (\ref{otocat4}). Note, however, that the approximation breaks down at large values of $K$ as the higher order terms in $K$ become significant and can no longer be ignored.

One can generalize this analysis to the other translationally invariant cases by imposing suitable conditions on $\omega$. For instance, an analogous condition for $R=3$ in the deep quantum regime reads $\omega=\sqrt{3}/4k\pi$, where $k\in\mathbb{Z}^+$, thereby demonstrating quadratic growth of the OTOC when observed at $t=3s$, with $s$ being a non-negative integer.
An upshot of this analysis is that the quantum resonances hinder the early-time exponential behavior, irrespective of the choice of other free parameters, such as the kicking strength $K$. It is worth noting that in finite-dimensional chaotic systems (or systems with torus boundary conditions in the classical limit), the OTOCs typically saturate to values predicted by the random matrix theory \cite{GMata2023}. In contrast, finding an unbounded operator growth is often possible in infinite-dimensional systems like the KHO. Quantum resonances are one such example. Here, we leveraged their solvability to show the indefinite operator growth.

\subsection{Small $K$ and irrational $R$}
\label{irr-der}
In the last section, we argued that whenever $R$ takes irrational values, various terms in Eq. (\ref{bosonic-evolution1}) destructively interfere, given that $K$ is small. As a result, the corresponding dynamics are suppressed. Here, we extend this argument by providing an analytical expression for the average state-OTOC, $\overline{C_{|\psi\rangle,\thinspace \hat{a}\hat{a}^{\dagger}}}(t)$, in the weak perturbative regime by assuming irrational values for $R$. The complete derivation is presented in Appendix \ref{appendix:b}. Retaining only the leading order terms in $K$ up to the order of $O(K^2)$, the Heisenberg evolution of $\hat{a}$ can be written as follows:
\begin{align}
\hat{a}(t)e^{\frac{2\pi it}{R}}&\approx\hat{a}+\dfrac{iK}{\sqrt{2\omega}}\sum_{j=0}^{t-1}e^{\frac{2\pi ij}{R}}\sin\left(\hat{X}_{\frac{2\pi j}{R}}\right)\nonumber\\
&-\dfrac{K^2}{\sqrt{2\omega}}\sum_{j=0}^{t-1}\sum_{n=0}^{j-1}e^{\frac{2\pi ij}{R}}\left[\cos\left(\hat{X}_{\frac{2\pi n}{R}}\right),\thinspace \sin\left(\hat{X}_{\frac{2\pi j}{R}}\right) \right],
\end{align}
where $\hat{X}_{\theta}=(\hat{a}e^{-i\theta}+\hat{a}^{\dagger}e^{i\theta})/\sqrt{2\omega}$, the quadrature operator with the phase $\theta$. We have taken $\hbar=1$ for the sake of simplicity. In the next step, we plug this into Eq. (\ref{com}). This is then followed by the absolute squaring of the commutator --- $[\hat{a}(t), \hat{a}^{\dagger}]^{\dagger}[\hat{a}(t), \hat{a}^{\dagger}]$, which is given in Eq. (\ref{irr-comm-comp}) of Appendix \ref{appendix:b}. After averaging over the pure states, we finally obtain
\begin{figure}
\includegraphics[scale=0.35]{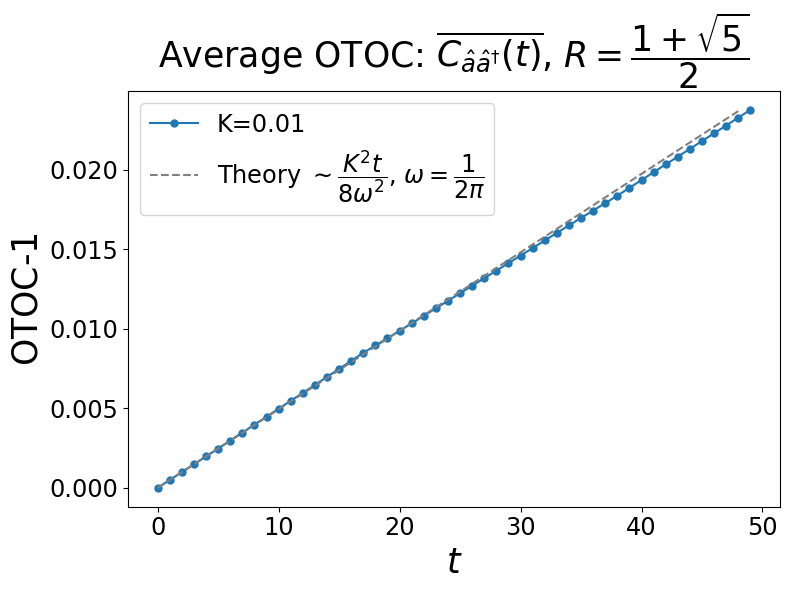}
\caption{\label{fig:otocatgm} The averaged commutator function $\overline{C_{\hat{a}\hat{a}^{\dagger}}(t)}$ in a weak perturbative regime when $R$ is irrational. We fix $R$ as the golden mean number, known to be the most irrational number. The figure compares the numerically computed average OTOC with the theoretical prediction given in Eq. (\ref{irr-otoc}). For the numerics, we average the OTOC over two thousand Fock states.}
\end{figure}
\begin{equation}\label{irr-otoc}
\overline{C_{|\psi\rangle,\thinspace \hat{a}\hat{a}^{\dagger}}}(t)\approx 1+\dfrac{K^2 t}{8\omega^2},\hspace{0.5cm}\text{for }K\ll 1
\end{equation}
To test Eq. (\ref{irr-otoc}), we numerically evaluate the commutator function averaged over the Fock state basis when $R$ assumes an irrational number. For the numerical calculations, we take $R$ to be the golden mean number --- $R=(1+\sqrt{5})/2$. Figure. \ref{fig:otocatgm} compares the numerical result with Eq. (\ref{irr-otoc}). We see a good agreement between the both.

We assert that the above result is not limited to the KHO model under the irrationality of $R$ but rather encompasses a broader range of KAM systems, including that of the finite dimensions. To be precise, the linear growth here is a direct consequence of (i) the uncorrelated eigenphases of the unperturbed evolution $e^{-(2\pi i/R)\hat{a}^{\dagger}\hat{a}}$, and (ii) the initial operators being conserved up to a phase under the unperturbed evolution, i.e., $e^{(2\pi i/R)\hat{a}^{\dagger}\hat{a}}\hat{a}e^{-(2\pi i/R)\hat{a}^{\dagger}\hat{a}}=\hat{a}e^{-2\pi i/R}$. One can readily verify that the average state-OTOC in a typical integrable quantum system perturbed by a weak generic time-dependent potential, in general, exhibits linear growth until the saturation as long as the above two conditions are satisfied --- $\overline{C(t)}\sim \varepsilon^2t$ with $\varepsilon$ being the strength of the perturbation. Let us also mention that, in general, the first condition holds for any typical KAM integrable system owing to the Berry-Tabor conjecture \cite{Berry375, Berry77a, berry1976closed}. In Appendix \ref{appendix:d}, we provide a detailed derivation for the linear growth of the OTOCs in finite-dimensional quantum systems by incorporating the aforementioned conditions.

\section{OTOC for phase space operators}
\label{OTOCXP}
For completeness, we here provide a bird's-eye view of the position-momentum OTOC in an arbitrary state $|\psi\rangle$ given by
\begin{eqnarray}
C_{XP}(t)=\langle\psi|\left[\hat{X}(t), \hat{P}\right]^{\dagger}\left[\hat{X}(t), \hat{P}\right]|\psi\rangle. 
\end{eqnarray}
Followed by Eq. (\ref{bosonic-evolution1}), the Heisenberg evolution of $\hat{X}$ can be readily obtained as 
\begin{equation}\label{Xevol}
\hat{X}(t)=u^{\dagger t}\hat{X}u^t +\dfrac{K}{\omega}\sum_{j=0}^{t-1}\sin\left[\dfrac{2\pi}{R}\left(t-j\right)\right]\sin\hat{X}(j),
\end{equation}
where $u=e^{(2\pi it/R) \hat{a}^{\dagger}\hat{a}}$, the evolution under the harmonic oscillator Hamiltonian with an effective frequency $\omega\tau=2\pi/R$. Due to the presence of $u$, the OTOC exhibits persistent oscillations in time. On the other hand, the second term takes into account the effects of the kicking potential on operator growth.
\begin{figure}
\includegraphics[scale=0.35]{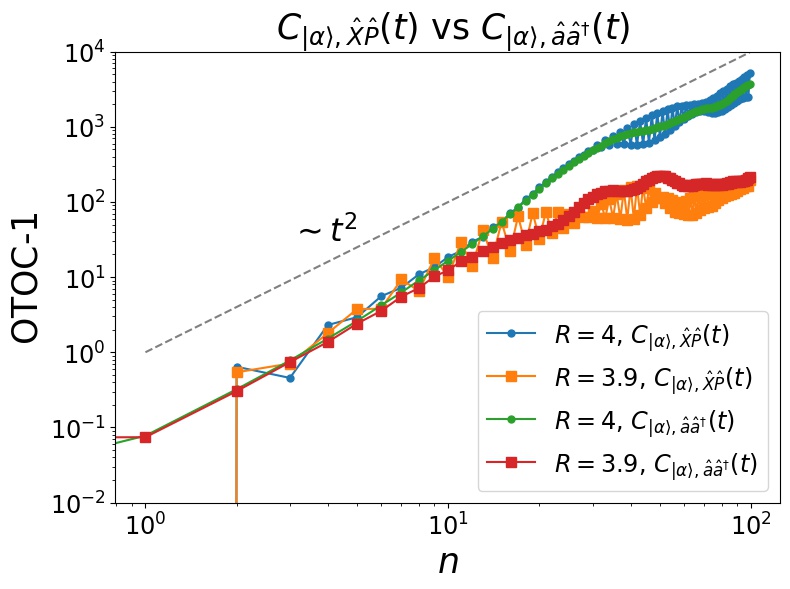}
\caption{\label{fig:xpotoc} Contrasting the OTOCs corresponding to the phase space operators with the ladder operators in the vicinity of $R=4$. The system parameters are kept fixed at $K=1$ and $\tau=1$. The coherent state centered at $(\hat{P}, \hat{X})=(0, \pi)$ is once again chosen to be the initial state for the OTOC calculations. The figure illustrates that the $\hat{X}\hat{P}$-OTOC and the $\hat{a}\hat{a}^{\dagger}$-OTOC exhibit strikingly similar behaviors in both the resonant and non-resonant cases, with the only difference being the presence of oscillations in the former.}
\end{figure}

When $K=0$, the OTOC in any given state $|\psi\rangle$ is $C_{|\psi\rangle,\thinspace \hat{X}\hat{P}}(t)=\hbar^2 \cos^2(2\pi t/R)$. As $K$ varies from zero, the initial operator $\hat{X}$ starts to scramble into the operator Hilbert space, which is also accompanied by the oscillatory behavior.
We numerically compute the $\hat{X}\hat{P}$-OTOC and compare the results with the ladder operator OTOC. The findings are shown in Fig. \ref{fig:xpotoc}. The figure demonstrates that the qualitative nature of the OTOCs for both choices of the initial operators is identical, except for the presence of oscillations in the $\hat{X}\hat{P}$-OTOC due to the harmonic evolution. These oscillations may pose challenges while probing the operator growth and quantum Lyapunov exponents. Nevertheless, a comparison between Equation (\ref{bosonic-evolution1}) and Equation (\ref{Xevol}) reveals that the oscillating component is absent in the time evolved ladder operator $\hat{a}(t)$. As a result, we observe steady growth without oscillations in the ladder operator OTOC, prompting us to study it as an alternative to the $\hat{X}\hat{P}$-OTOC.

\section{Summary and Discussion}
\label{section-4}


In this work, we have studied how the resonances that arise when a degenerate classical system is perturbed by a weak time-dependent potential affect the dynamics of information scrambling in the quantum domain. For this purpose, we considered the kicked harmonic oscillator model. Classically, this system exhibits very complex dynamics. Under the resonance condition, the system undergoes unusual structural changes in the phase space. These unusual changes can be traced to the emergence of stable and unstable periodic orbits, leading to diffusive chaos in the phase space for any finite kicking strength. Conversely, under the non-resonance condition, the phase space trajectories of the harmonic oscillator maintain their regularity with slight deformations, thereby effectively suppressing the chaos. Motivated by this peculiar classical behavior, our work examined how these effects manifest in the dynamics of operator growth and scrambling in the quantum limit.     

To study the information scrambling, we considered the OTOC with the bosonic ladder operators as the initial operators. We mainly focussed on two specific initial states: the vacuum state $|0\rangle$ that corresponds to a fixed point of the KHO map and a coherent state $|\alpha\rangle$ with its mean coordinates located on the stochastic web. We numerically studied the OTOC, particularly emphasizing the early-time and the asymptotic dynamics. In the semiclassical limit, the early-time dynamics correlate very well with the classical dynamics of the KHO. In this region, the quantum Lyapunov exponents extracted from the OTOC match very well with the corresponding classical Lyapunov exponents. To strengthen the correspondence between the classical and quantum exponents, we analytically derived the quantum exponent from the vacuum state OTOC in the semiclassical limit. Furthermore, we observed that in the weak perturbative regime, the transitions between resonance and non-resonance conditions do not affect the early-time growth, while the asymptotic dynamics remain highly sensitive. On the contrary, when the perturbation is strong, the differences become less visible. 

The numerical and analytical evidence presented in the paper suggests a long-time quadratic growth for the OTOC when the system is in resonance for a generic quantum state irrespective of the strength of the perturbation. In contrast, the growth under the non-resonance condition is largely suppressed. We also identified different scalings for the non-resonant OTOCs in the coherent state. To support the numerical results, we have argued based on the coherent and incoherent additions of the terms in the expansion of $[\hat{a}(t), \hat{a}^{\dagger}]e^{2\pi it/R}$ that demonstrate the distinct qualitative behavior exhibited by the resonances and the non-resonances. 

Following the numerical results, we provided analytical treatment of the OTOC for a few exceptional cases of $R\in R_c$ by utilizing the translation invariance of the KHO. For $R=1$ and $2$, the results are exact. We have shown that the corresponding commutator function grows quadratically. At $R=4$, we utilized the quantum resonance condition to obtain the quasi-exact expressions of the OTOCs. The quantum resonances largely impede the early-time exponential growth of the OTOCs. We also provided analytical derivation for the linear growth of the average state-OTOC whenever $R$ takes irrational values, given that the kicking strength is sufficiently small. 

One central focus in quantum many-body physics is to simulate quantum systems on a quantum device and benchmark these simulations in the presence of hardware errors \cite{sahu2022quantum, trivedi2022quantum}. In the semiclassical limit, the assurance will be provided for the stability of simulation if the classical counterpart of the target Hamiltonian is KAM and the error that scales extensively with the number of particles is small enough \cite{bulchandani2022onset}. However, for systems that have no classical analogue, the quantum KAM theorem remains elusive. Nevertheless, recent progress in this direction studied the stability analysis for the symmetries of quantum systems --- see for instance, Refs. \cite{brandino2015glimmers, burgarth2021kolmogorov} and the references therein. Quantum simulations of systems whose classical limit is non-KAM can pose a significant challenge to experimental implementations. Any slight perturbation in the form of hardware noise can give rise to dynamics far from the target dynamics. Moreover, digital quantum simulation of these systems is also challenging due to structural changes near resonances even when the hardware error is negligible \cite{chinni2022trotter}. Hence, more careful methods must be devised to simulate this class of systems. We hope our study paves the way for exploring these intriguing directions.   

\vspace{.0cm}
\begin{acknowledgments}
The authors acknowledge funding from the Department of Science and Technology, Govt of India, under Grant No. DST/ICPS/QusT/Theme-3/2019/Q69 and New faculty Seed Grant from IIT Madras. This work was supported, in part, by a grant from Mphasis to the Centre for Quantum Information, Communication, and Computing (CQuICC) at IIT Madras.
\end{acknowledgments}

\bibliographystyle{unsrt}
\bibliography{nonKAM}
\onecolumngrid
\appendix

\section{Computation of $\overline{\langle D(\beta)\rangle }$}\label{appendix:a}
In this appendix, we find the average of the expectation value of an arbitrary displacement operator over the space of all pure states, which facilitates the derivations of the average state-OTOCs discussed in the main text. Consider a displacement operator $D(\beta),$ where $\beta\in\mathbb{C}$. We are interested in evaluating $\overline{\langle D(\beta)\rangle}=\int_{\psi}d\psi\langle\psi|D(\beta)|\psi\rangle$, where $d\psi$ represents the normalized uniform measure over the space of pure states in the infinite-dimensional Hilbert space. Since the Fock states of the quantum harmonic oscillator form an orthonormal basis and constitute continuous variable state 1-designs, it suffices to average $\langle D(\beta)\rangle$ over the set of Fock states.
\begin{eqnarray}\label{dispavg}
\overline{\langle D(\beta)\rangle }=\int_{\psi}d\psi\langle\psi|D(\beta)|\psi\rangle&\equiv&\lim_{N\rightarrow \infty}\dfrac{1}{N} \sum_{n=0}^{N-1}\langle n|D(\beta)|n\rangle
\end{eqnarray}
The elements of the displacement operator in the Fock state basis can be written in terms of the associated Laguerre polynomials. Specifically, the diagonal entries, $\langle n|D(\beta)|n\rangle$ for all $n\geq 0$, can be obtained as 
\begin{equation}
\langle n|D(\beta)|n\rangle=e^{-|\beta|^2/2}L^0_n(|\beta|^2), \text{ where }L^0_n\left(|\beta|^2\right)=\sum_{k=0}^{n}(-1)^k\dfrac{n!}{(n-k)!k!k!}|\beta|^{2k}. 
\end{equation}
After incorporating this into Eq. (\ref{dispavg}), we get
\begin{eqnarray}\label{dispavgfinal}
\overline{\langle D(\beta)\rangle }&=&\lim_{N\rightarrow\infty}\dfrac{1}{N}\sum_{n=0}^{N-1}e^{-|\beta|^2/2}L^0_{n}(|\beta|^2)\nonumber\\
&=&e^{|\beta|^2/2}\lim_{N\rightarrow\infty}\dfrac{L^1_{N-1}(|\beta|^2)}{N} \nonumber\\
&=&e^{|\beta|^2/2}\delta_{\Re\left(\beta\right), 0}\delta_{\Im\left(\beta\right), 0}\nonumber\\
&=&\delta_{\Re\left(\beta\right), 0}\delta_{\Im\left(\beta\right), 0}.
\end{eqnarray}
In the second equality, we used the recursive relation $\sum_{n=0}^{k}L^i_{n}(x)=L^{i+1}_{k}$. Therefore, the average vanishes unless $\beta$ is zero, in which case $D(\beta)$ reduces to an infinite dimensional identity operator.

\section{Small $K$ Approximation of OTOCs}
\label{appendix:b}
In this appendix, we aim to achieve two objectives. First, we provide analytical arguments to demonstrate that for small $K$, the commutator function exhibits quadratic growth at resonances. We then give an explicit derivation for Eq. (\ref{irr-otoc}) --- the average state-OTOC for irrational $R$ and small $K$. Here, we take $\hbar=1$. The derivation involves approximating the commutator $[\hat{a}(t), \hat{a}]$ up to the terms of order $O(K^2)$. 
\begin{eqnarray*}
[\hat{a}(t), \hat{a}]e^{2\pi it/R}=1+\dfrac{iK}{\sqrt{2\omega}}\sum_{j=0}^{t-1}e^{2\pi ij/R}\left[\sin\hat{X}(j), \hat{a}^{\dagger}\right], \text{ where } X=\dfrac{\hat{a}+\hat{a}^{\dagger}}{\sqrt{2\omega}}.
\end{eqnarray*}
To approximate the commutator up to the second order in $K$, we only require to retain $\sin\hat{X}(j)$ to the zeroth and the first order terms in $K$ for all $j>0$. First, a single application of $\hat{U}_{\tau}$ on $\sin\hat{X}$ gives
\begin{eqnarray}
\sin\hat{X}(1)&=&e^{iK\cos\hat{X}}e^{i\omega\tau\hat{a}^{\dagger}\hat{a}}\left(\sin\hat{X}\right)e^{-i\omega\tau\hat{a}^{\dagger}\hat{a}}e^{-iK\cos\hat{X}}\nonumber\\
&=&\sin\left(\hat{X}_{2\pi /R}\right)+iK\left[\cos\hat{X}, \sin\left(\hat{X}_{2\pi/R}\right)\right]+O(K^2),
\end{eqnarray}
where $\hat{X}_{\theta}=(\hat{a}e^{-i\theta}+\hat{a}^{\dagger}e^{i\theta})/\sqrt{2\omega}$. After $j$-number of repeated applications, the time evolved operator $\sin\hat{X}(j)$ can be approximated as shown below:
\begin{equation}
\sin\hat{X}(j)= \sin\left(\hat{X}_{2\pi j/R}\right)+iK\sum_{n=0}^{j-1}\left[\cos\left(\hat{X}_{2\pi n/R}\right), \sin\left(\hat{X}_{2\pi j/R}\right)\right]+O(K^2).
\end{equation}
Consequently, the Heisenberg evolution of $\hat{a}$ can be approximated as
\begin{equation}
\hat{a}(t)\approx \hat{a}+\dfrac{iK}{\sqrt{2\omega}}\sum_{j=0}^{t-1}e^{ij\omega\tau}\left\{\sin\left(\hat{X}_{2\pi j/R}\right)+iK\sum_{n=0}^{j-1}\left[\cos\left(\hat{X}_{2\pi n/R}\right),\sin\left(\hat{X}_{2\pi j/R}\right) \right]\right\}.
\end{equation}
From Eq. (\ref{com}), it follows that
\begin{eqnarray}\label{com-app}
\left[\hat{a}(t), \hat{a}^{\dagger}\right]e^{2\pi it/R}&\approx &1+\dfrac{K}{2\omega}\sum_{j=0}^{t-1}\cos\left(\hat{X}_{2\pi j/R}\right)-\dfrac{K^2}{\sqrt{2\omega}}\sum_{j=0}^{t-1}\sum_{n=0}^{j-1}e^{2\pi ij/R}\left[\left[\cos\left(\hat{X}_{2\pi n/R}\right),\sin\left(\hat{X}_{2\pi j/R}\right) \right], \hat{a}^{\dagger}\right]
\end{eqnarray}
This approximation is valid for any $R$ as long as $K$ is small. In the following, we will focus on two separate instances: $R=4$ and an irrational value of $R$.

\textit{Instance-1 :} 
In the main text, we argued that the resonances usually result in coherent summations, leading to the quadratic growth of OTOCs. To illustrate this further, we examine Eq. (\ref{com-app}) for the case of $R=4$. To simplify the computation, we consider $t=4s$ where $s$ is a non-negative integer. By doing so, we can observe that the first summation on the right-hand side exhibits a clear linear dependence on $t$.
\begin{eqnarray}
\sum_{j=0}^{t-1}\cos\left(\hat{X}_{2\pi j/R}\right)= \dfrac{t}{4}\sum_{j=0}^{3}\cos\left(\hat{X}_{2\pi j/R}\right). 
\end{eqnarray}
on the other hand, the third term that contains the double summations has an implicit quadratic time dependence. Therefore, in the limit of weak perturbations, the OTOCs will grow quadratically at resonances. The proof of the argument is now complete.

\textit{Instance-2: }
We now consider $R$ to be an irrational number. The approximate OTOC in the limit of small $K$ can be written as
\begin{eqnarray}\label{irr-comm-comp}
C_{\hat{a}\hat{a}^{\dagger}}(t)&\approx & 1+\dfrac{K^2}{4\omega^2}\sum_{j, j'=0}^{t-1}\cos\left(\hat{X}_{2\pi j/R}\right)\cos\left(\hat{X}_{2\pi j'/R}\right)-\dfrac{K^2}{\sqrt{2\omega}}\sum_{j=0}^{t-1}\sum_{n=0}^{j-1}\left\{e^{2\pi ij/R}\left[\left[\cos\left(\hat{X}_{2\pi n/R}\right),\sin\left(\hat{X}_{2\pi j/R}\right) \right], \hat{a}^{\dagger}\right]+\textbf{h.c.}\right\}
\end{eqnarray}
Upon performing the average over the pure states, the third term vanishes for all $j<t$ and $n<j$. The second term also vanishes unless $j= j'$. Therefore, we finally obtain
\begin{eqnarray}
\overline{C_{|\psi\rangle,\thinspace \hat{a}\hat{a}^{\dagger}}}(t)&\approx &1+\dfrac{K^2}{4\omega^2}\sum_{j=0}^{t-1}\overline{\langle\psi| \cos^2\left(\hat{X}_{2\pi j/R}\right)|\psi\rangle}\nonumber\\
&=&1+\dfrac{K^2t}{8\omega^2} \text{ for } K\ll 1.
\end{eqnarray}
This concludes the derivation of Eq. (\ref{irr-otoc}) in the main text. 

\section{Linear growth of OTOCs in finite-dimensional integrable quantum systems}
\label{appendix:d}
It is well-known that a generic quantum system, whose classical limit is integrable, possesses uncorrelated eigenspectrum (or eigenphases if the system is time-periodic). If $V$ denotes the time evolution of a typical integrable system, then its eigenphases can be viewed as complex phases drawn uniformly at random from a complex unit circle. Without further loss of generality, we assume $V$ to be a random diagonal unitary acting on a $d$-dimensional Hilbert space. The elements of $V$ can be characterized as follows:
\begin{eqnarray}
V_{i j}=
\begin{cases}
 e^{2\pi i\phi}, \phi\in \left[-0.5, 0.5\right]& \text{if }  i=j\\
    0,              & \text{otherwise},
\end{cases}
\end{eqnarray}
where $\phi$ is a uniform random variable. After applying the perturbation, we assume the following Floquet form for the total system evolution: 
\begin{eqnarray}
U=Ve^{-i\varepsilon H} \text{ and }\varepsilon\ll 1, 
\end{eqnarray}
where $H$ is the perturbation and $\varepsilon$ is the kicking strength. By choosing the uncorrelated eigenphases, we already invoked the first condition that led to the linear growth of the OTOCs under the non-resonance condition. We now choose the initial operators that are conserved up to a phase under the action of $V$ --- let $A$ be an operator that satisfies $V^{\dagger}AV=e^{-i\theta}A$. For simplicity, we take $\theta=0$. We are now interested in computing the quantity given by 
\begin{eqnarray}
C(t)=\dfrac{1}{d}\text{Tr}\left[\left[A(t), A\right]^{\dagger}\left[A(t), A\right]\right], 
\end{eqnarray}
where $A(t)=U^{\dagger t}AU^t$. Ignoring all the higher order terms in $\varepsilon$, the commutator, $\left[A(t), A\right]$, can be written as 
\begin{eqnarray}
\left[A(t), A\right]\approx i\varepsilon\sum_{j=0}^{t-1}\left[\left[V^{\dagger j}HV^j, A\right], A\right].
\end{eqnarray}
It then follows that
\begin{eqnarray}
\text{Tr}\left[\left[A(t), A\right]^{\dagger}\left[A(t), A\right]\right] &\approx&\varepsilon^2\sum_{j, j'=0}^{t-1}\left[\left[V^{\dagger j}HV^j, A\right], A\right]^\dagger\left[\left[V^{\dagger j'}HV^{j'}, A\right], A\right]\nonumber\\
&=&\varepsilon^2\sum_{j, j'=0}^{t-1}\text{Tr}\left[6H_jA^2H_{j'}A^2-4H_jA^3H_{j'}A-4H_jAH_{j'}A^3+H_jH_{j'}A^4+H_{j'}H_jA^4\right]\nonumber\\
&=&\varepsilon^2\sum_{j, j'=0}^{t-1}\text{Tr}\left[6H_{j-j'}A^2HA^2-4H_{j-j'}A^3HA-4H_{j-j'}AHA^3+H_{j-j'}HA^4+HH_{j-j'}A^4\right]
\end{eqnarray}
where $H_{j-j'}=V^{\dagger j-j'}HV^{j-j'}$. Now, the process of averaging over the random diagonal unitaries gives
\begin{eqnarray}\label{C6}
\int_{V}dVC(t)&=&\dfrac{1}{d} \int_{V}\text{Tr}\left[\left[A(t), A\right]^{\dagger}\left[A(t), A\right]\right]dV\nonumber\\
&=&\dfrac{\varepsilon^2}{d}\sum_{j, j'=0}^{t-1}\int_{V}dV\text{Tr}\left[6H_{j-j'}A^2HA^2-4H_{j-j'}A^3HA-4H_{j-j'}AHA^3+H_{j-j'}HA^4+HH_{j-j'}A^4\right]\nonumber\\
&=&\dfrac{\varepsilon^2}{d}t\text{Tr}\left[6HA^2HA^2+2H^2A^4-8HAHA^3\right]\nonumber\\
&&+\dfrac{\varepsilon^2}{d}\sum_{j\neq j'}\int_{V}dV\text{Tr}\left[6H_{j-j'}A^2HA^2-4H_{j-j'}A^3HA-4H_{j-j'}AHA^3+H_{j-j'}HA^4+HH_{j-j'}A^4\right].
\end{eqnarray}
The integrals over the random diagonal unitaries can be solved as follows:
\begin{eqnarray}
\int_{V}dV \text{ Tr}\left[H_{j-j'}A^2HA^2\right]&=&\text{ Tr}\left[\text{diag}\left(H\right)\right]\text{Tr}\left[HA^4\right],\nonumber\\
\int_{V}dV\text{ Tr}\left[H_{j-j'}A^3HA\right]&=&\text{ Tr}\left[\text{diag}\left(H\right)\right]\text{Tr}\left[HA^4\right],\nonumber\\
\int_{V}dV\text{ Tr}\left[H_{j-j'}HA^4\right]&=&\text{ Tr}\left[\text{diag}\left(H\right)\right]\text{Tr}\left[HA^4\right], \nonumber\\
\int_{V}dV\text{ Tr}\left[HH_{j-j'}A^4\right]&=&\text{ Tr}\left[\text{diag}\left(H\right)\right]\text{Tr}\left[HA^4\right].\nonumber\\
\end{eqnarray}
All the above integrals yield identical results. As a result, the second term in Eq. (\ref{C6}) involving the double summation vanishes. Therefore, we finally obtain 
\begin{eqnarray}
\int_{V}dVC(t)=\dfrac{\varepsilon^2 t}{d}\text{ Tr}\left[6HA^2HA^2+2H^2A^4-8HAHA^3\right]
\end{eqnarray}
In conclusion, we have shown analytically that the OTOC grows linearly with time, given that the aforementioned conditions regarding the initial operators and eigenphases are satisfied.
\end{document}